\documentclass[aps,amsmath,preprint,epsf,superscriptaddress,nofootinbib]{revtex4-1}
\usepackage{graphicx}
\usepackage{bm}
\usepackage{color}
\usepackage{amsmath,amssymb}	
\usepackage{hyperref}
\usepackage{setspace}
\usepackage{longtable}
\usepackage{booktabs}
\usepackage{comment}
\usepackage{array}
\usepackage{cancel}
\usepackage{enumitem}

\def\slashchar#1{\setbox0=\hbox{$#1$} % set a box for #1
\dimen0=\wd0 % and get its size
\setbox1=\hbox{/} \dimen1=\wd1 % get size of /
\ifdim\dimen0>\dimen1 % #1 is bigger
\rlap{\hbox to \dimen0{\hfil/\hfil}} % so center / in box
#1 % and print #1
\else % / is bigger
\rlap{\hbox to \dimen1{\hfil$#1$\hfil}} % so center #1
/ % and print /
\fi}

\def\q{\partial}

\def\D{\Delta}

\def\beq{\begin{eqnarray}}
\def\eeq{\end{eqnarray}}

\newcommand{\vev}[1]{ \left\langle {#1} \right\rangle }

\begin{document}
\newcolumntype{Y}{>{\centering\arraybackslash}p{23pt}} 

%%%%%%%%%%%%%%%%%%%%%%%%%%%%%%%%%%
%%%%%%%%%%% Title page %%%%%%%%%%%
%%%%%%%%%%%%%%%%%%%%%%%%%%%%%%%%%%

\preprint{IPMU18-0095}

\title{\textit{B}\,--\,\textit{L} as a Gauged Peccei-Quinn Symmetry}

\author{Masahiro Ibe}
\email[e-mail: ]{ibe@icrr.u-tokyo.ac.jp}
\affiliation{Kavli IPMU (WPI), UTIAS, The University of Tokyo, Kashiwa, Chiba 277-8583, Japan}
\affiliation{ICRR, The University of Tokyo, Kashiwa, Chiba 277-8582, Japan}
\author{Motoo Suzuki}
\email[e-mail: ]{m0t@icrr.u-tokyo.ac.jp}
\affiliation{Kavli IPMU (WPI), UTIAS, The University of Tokyo, Kashiwa, Chiba 277-8583, Japan}
\affiliation{ICRR, The University of Tokyo, Kashiwa, Chiba 277-8582, Japan}
\author{Tsutomu T. Yanagida}
\email[Hamamatsu Professor. e-mail: ]{tsutomu.tyanagida@ipmu.jp}
\affiliation{Kavli IPMU (WPI), UTIAS, The University of Tokyo, Kashiwa, Chiba 277-8583, Japan}

\date{\today}
\begin{abstract}
The gauged Peccei-Quinn (PQ) mechanism provides a simple prescription to embed the global PQ symmetry into a gauged $U(1)$ symmetry.
As it originates from the gauged PQ symmetry, the global PQ symmetry can be protected from explicit breaking by quantum gravitational effects
once appropriate charge assignment is given.
In this paper, we identify the gauged PQ symmetry with the ${B-L}$ symmetry,
which is obviously attractive as the ${B-L}$ gauge symmetry is the most authentic extension of the Standard Model. 
As we will show, a natural $B-L$ charge assignment can be found in a model motivated by 
the seesaw mechanism in the $SU(5)$ Grand Unified Theory.
As a notable feature of this model, it does not require extra $SU(5)$ singlet matter fields 
other than the right-handed neutrinos to cancel the self and the gravitational anomalies.
\end{abstract}

\maketitle

\section{Introduction}
\label{sec:introduction}
%PQ 
The strong CP problem is longstanding and probably one of the most puzzling issues in particle physics.
Although the Peccei-Quinn (PQ) mechanism\,\cite{Peccei:1977hh,Peccei:1977ur,Weinberg:1977ma,Wilczek:1977pj} 
provides a successful solution to the problem, it is not very satisfactory from a theoretical point of view, 
as it  relies on a global Peccei-Quinn $U(1)$ symmetry.
The Peccei-Quinn symmetry is required to be almost exact but explicitly broken by the QCD anomaly.
Even tiny explicit breaking terms of the PQ symmetry spoil the PQ mechanism.
It is, on the other hand, conceived that any global symmetries are broken by quantum gravity 
effects~\cite{Hawking:1987mz,Lavrelashvili:1987jg,Giddings:1988cx,Coleman:1988tj,Gilbert:1989nq,Banks:2010zn}.
Thus, the PQ mechanism brings up another question, the existence of such an almost but not exact global symmetry.

In \cite{Fukuda:2017ylt},  a general prescription to achieve a desirable PQ symmetry 
is proposed in which the PQ symmetry originates from a gauged $U(1)$ symmetry, $U(1)_{gPQ}$.
The anomalies of $U(1)_{gPQ}$ are canceled between the contributions from two (or more) PQ charged sectors,
while the inter-sector interactions between the PQ charged sectors are highly suppressed by appropriate $U(1)_{gPQ}$ charge assignment.
As a result of the separation, a global PQ symmetry exist in addition to $U(1)_{gPQ}$ as an accidental symmetry.
The accidental PQ symmetry is highly protected from explicit breaking by quantum gravitational effects as it originates from the gauge symmetry.
The gauged PQ mechanism is a generalization of the mechanisms which achieve the PQ symmetry 
as an accidental symmetry resulting from (discrete) gauge 
symmetries~\cite{Barr:1992qq,Kamionkowski:1992mf,Holman:1992us,Dine:1992vx,Dias:2002gg,
Carpenter:2009zs,Harigaya:2013vja,Dias:2014osa,Ringwald:2015dsf,Harigaya:2015soa,Redi:2016esr,Duerr:2017amf}.

In this paper, we discuss whether the $B-L$ symmetry can play a role of the gauged PQ symmetry.
The ${B-L}$ gauge symmetry is the most authentic extension of the Standard Model (SM) which 
explains the tiny neutrino masses via the seesaw mechanism~\cite{Yanagida:1979as,GellMann:1980vs,Glashow:1979nm} (see also \cite{Minkowski:1977sc}). 
Therefore, the identification of the gauged PQ symmetry with  $B-L$ makes the gauged PQ mechanism more plausible.\
An intriguing coincidence between  the right-handed neutrino mass scale appropriate for the thermal leptogenesis~\cite{Fukugita:1986hr}~(see \cite{Giudice:2003jh,Buchmuller:2005eh,Davidson:2008bu}, for review) and the PQ breaking scale which avoids astrophysical constraints 
also motivates this identification~\cite{Langacker:1986rj}.

As will be shown, we find a natural $B-L$ charge assignment motivated by the seesaw mechanism in the $SU(5)$ Grand Unified Theory (GUT),
with which the gauged PQ mechanism is achieved.
Notably, the charge assignment we find does not require extra $SU(5)$ singlet matter fields other than the right-handed neutrinos
to cancel the $[U(1)_{gPQ}]^3$ and the gravitational anomalies.

The organization of the paper is as follows.
In section~\ref{sec:B-L}, we discuss an appropriate $B-L$ charge assignment so that it plays a role of $U(1)_{gPQ}$.
In section~\ref{sec:axion}, we discuss the properties of the axion and the global PQ symmetry.
In section~\ref{sec:DW}, we  briefly discuss the domain wall problem.
In section~\ref{sec:susy}, we discuss supersymetric (SUSY) extension of the model.
 The final section is devoted to our conclusions.

%%%%%%%%%%%%%%%%%%%%%%%%%%%%%%%%%%%%%%%%%%
\section{\textit{B}\,--\,\textit{L} as a gauged PQ symmetry}
\label{sec:B-L}
Among the various extension of the Standard Model, $B-L$ is the most plausible addition.
The anomalies of the $B-L$ gauge symmetry are canceled by simply introducing three SM singlet right-handed neutrinos $\bar{N}_R$.
The $B-L$ extended Standard Model naturally implements the seesaw mechanism by the spontaneous breaking of $B-L$  at the 
intermediate scale.

Having the $SU(5)$ GUT in mind,
it is more convenient to consider  ``fiveness", $5(B-L) -4Y$, instead of $B-L$, 
as it commutes with the $SU(5)$ gauge group.
The fiveness charges of the matter fields are given by
\begin{eqnarray}
\label{eq:SMfiveness}
{\mathbf{10}}_{\rm SM} (+1)\ , 
\quad
{\mathbf {\bar 5}}_{\rm SM} (-3)\ , 
\quad
\bar{N}_R(+5)\ ,
\end{eqnarray}
while the Higgs doublet, $h$, has a charge $+2$ (i.e. $B-L = 0$).%
\footnote{The colored Higgs is assumed to obtain a mass of the GUT scale.}
Here, we use the $SU(5)$ GUT representations for the matter fields, i.e. ${\mathbf{10}}_{\rm SM} $ = ($q_L,\bar{u}_R,\bar{e}_R$) 
and ${\mathbf {\bar 5}}_{\rm SM}$  = ($\bar{d}_R,\ell_L$), while $\bar{N}_R$ denotes the right-handed neutrinos.

The seesaw mechanism is implemented by assuming that the right-handed neutrinos obtain Majorana masses from 
spontaneous breaking of fiveness.
In this paper, we assume that the Majorana masses are provided by the vacuum expectation value (VEV) of a gauge singlet 
scalar field with fiveness, $-10$, i.e., 
\begin{eqnarray}
\phi(-10)\ ,
\end{eqnarray}
which couples to the right handed neutrinos,
\begin{eqnarray}
{\cal L} = - \frac{1}{2} y_N \phi \bar{N}_{R}\bar{N}_{R} + h.c. \ .
\end{eqnarray}
Here, $y_N$ denotes a coupling constant, with which the Majorana mass is given by $M_N = y_N\vev{\phi}$.
By integrating out the right-handed neutrinos, the tiny neutrino masses are obtained, via
\begin{eqnarray}
{\cal L} = y_\ell {\mathbf{\bar 5}_{\rm SM}} \bar{N}_R h^* + h.c.  ,
\end{eqnarray}
where $y_\ell$ also denotes a coupling constant.

%%%%%%%%%%%%%%%%%%%%%%%%%%%%%%%%%%%%%%%%%%
Now, let us identify the gauged PQ symmetry with  $B-L$, i.e., fiveness.
Following the general prescription of the gauged PQ mechanism in \cite{Fukuda:2017ylt},
let us introduce extra matter multiplets which obtain a mass from the VEV of $\phi$;
\begin{eqnarray}
\label{eq:KSVZ1}
{\cal L} = y_K \phi^*\, {\mathbf{5}}_K {\mathbf{\bar 5}}_K + h.c. \ ,
\end{eqnarray}
with $y_K$ being a coupling constant.%
\footnote{The reason why the extra multiplets couple not to $\phi$ but  $\phi^*$ will become clear shortly.}
Here, the extra multiplets (${\mathbf{5}}_K$,${\mathbf{\bar 5}_K}$) are assumed to form the ${\mathbf 5}$ and ${\mathbf{\bar 5}}$ representations 
of the $SU(5)$ gauge group, respectively.
As in the KSVZ axion model~\cite{Kim:1979if, Shifman:1979if}, the Ward identity of the fiveness current, 
$j_{5}$, obtains an anomalous contribution from the extra multiplets,
\begin{eqnarray}
\label{eq:anom}
\q j_{5}\big|_{\rm SM+N+ K} =-\frac{g_a^2}{32\pi^2}10 F^a \tilde F^a\ .
\end{eqnarray}
Here, $F^a$ ($a$ = $1,2,3$) are the gauge field strengths of the Standard Model and $g_a$ the corresponding SM gauge coupling constants. 
The Lorentz indices and the gauge group representation indices are suppressed. 
The factor $-10$ corresponds to the charge of the bi-linear, ${\mathbf{5}}_K {\mathbf{\bar 5}}_K$ (see Eq.\,(\ref{eq:KSVZ1})).

In the gauged PQ mechanism, the $U(1)_{gPQ}$ gauge anomalies are canceled by a contribution from another set of 
the PQ charged sector.
For that purpose, let us also introduce $10$-flavors of extra  matter multiplets $( {\mathbf{5}}_K', {\mathbf{\bar 5}}_K' )$.
We assume that they obtain masses from a VEV of a complex scalar field $\phi'$ whose fiveness charge is $+1$;
\begin{eqnarray}
\label{eq:KSVZ2}
{\cal L} = y_K' \phi'^*\, {\mathbf{5}}_K' {\mathbf{\bar 5}}_K' + h.c. \ ,
\end{eqnarray}
where the charge of the bi-linear, ${\mathbf{5}}_K' {\mathbf{\bar 5}}_K'$, is set to be $+1$.
With this choice, the anomalous contributions of the Ward identity in (\ref{eq:anom}) are 
canceled by the one from $( {\mathbf{5}}_K', {\mathbf{\bar 5}}_K' )$, i.e.,
\begin{eqnarray}
\q j_{5}\big|_{\rm SM+N+ K + K'} = 0\ .
\end{eqnarray}

The fiveness charges of the respective extra multiplets are chosen as follows.
To avoid stable extra matter fields, we assume that  $\mathbf{\bar 5}_{K}$ and $\mathbf{\bar 5}_{K}'$
can mix with $\mathbf{\bar 5}_{\rm SM}$, so that
\begin{eqnarray}
\label{eq:charge}
\mathbf{5}_K(-7)\ , \,\, {\mathbf{\bar 5}}_K(-3) \ ,\,\, \mathbf{5}_K'(+4)\ ,\,\, {\mathbf{\bar 5}}_K'(-3) \ ,
\end{eqnarray}
respectively.
As a notable feature of this charge assignment, it cancels the $[U(1)_{gPQ}]^3$ and the gravitational anomalies
automatically without introducing additional SM singlet fields. 
In fact, the $[U(1)_{gPQ}]^3$ and the gravitational anomalies are proportional to
\begin{eqnarray}
[U(1)_{gPQ}]^3 &\propto& \left( (-10- \bar{q}_K)^3 + (\bar{q}_K)^3 \right) + 10\left( (1- \bar{q}_K')^3 + (\bar{q}_K')^3 \right) \ , \\
{}[\mbox{gravitational}] &\propto& \left( (-10- \bar{q}_K) + (\bar{q}_K) \right) + 10\left( (1- \bar{q}_K') + (\bar{q}_K') \right)\nonumber\ ,
\end{eqnarray}
with $\bar{q}_K$ and $\bar{q}_K'$ are the charges of $\mathbf{\bar 5}_K$ and $\mathbf{\bar 5}_K'$, respectively.
By substituting $\bar{q}_K = \bar{q}_K' = -3$, we find that both the anomalies are vanishing.%
%\footnote{Inversely, we find that the charge of $\phi'$ can be uniquely determined to be $-1$ if 
%we first fix $\bar{q}_K = \bar{q}_K' = -3$ with the charge of $\phi$ being $10$.}

The anomaly cancellation without  singlet fields other than the right-handed neutrinos is by far advantageous compared with 
the previous models~\cite{Barr:1992qq,Fukuda:2017ylt,Fukuda:2018oco}.
The singlet fields required for the anomaly cancellation tend to be rather light and longlived, which make the thermal history of the universe
complicated~\cite{Fukuda:2018oco}.
The anomaly cancellation of the present model is, therefore, a very important success as it is partly motivated by 
thermal leptogenesis which requires a high reheating temperature after inflation,
i.e., $T_R \gtrsim 10^{9}$\,GeV~\cite{Giudice:2003jh,Buchmuller:2005eh,Davidson:2008bu}.

Under the fiveness symmetry, the interactions are restricted to
\begin{eqnarray}
\label{eq:Lagrangian}
{\cal L} &=&  {\mathbf{10}_{\rm SM}}  {\mathbf{10}_{\rm SM}} h^*
+{\mathbf{10}_{\rm SM}}  {\mathbf{\bar{5}}}\, h
+ {\mathbf{\bar{5}}} \bar{N}_R h^*
- \frac{1}{2} \phi\, \bar{N}_{R}\bar{N}_{R} 
+ \phi^* \, \mathbf{5}_K \mathbf{\bar 5}+ \phi'{}^* \, \mathbf{5}_K' \mathbf{\bar 5} + h.c.\ ,\cr
&& - V(\phi,\phi',h) \ .
\end{eqnarray}
Here, $\mathbf{\bar 5}$ collectively denotes ($\mathbf{\bar 5}_{\rm SM}$,$\mathbf{\bar 5}_{K}$,$\mathbf{\bar 5}_{K}'$),
and $V(\phi, \phi',h)$ is the scalar potential.
The coupling coefficients are omitted for notational simplicity.
At the renormalizable level, the above Lagrangian possesses a global $U(1)$ symmetry, 
which is identified with the global PQ symmetry. 
The global PQ symmetry corresponds to a phase rotation of a gauge invariant combination, $\phi \phi'^{10}$,
while the other fields are rotated appropriately.
The global PQ charges of the individual fields are generically given by 
\begin{eqnarray}
\label{eq:PQQ}
Q = - \frac{Q_\phi}{10}\times q_5\ , \quad Q' =Q_{\phi'} -\frac{3}{10} Q_{\phi}\ ,
\end{eqnarray}
for $\{{\mbox{SM}}, \bar{N}_R, {\mathbf 5}_K, {\mathbf {\bar 5}}\}$
and $\{{\mathbf 5}_K'\}$, respectively.
Here, $q_5$ denotes the fivness charge of each field, and $Q_{\phi,\phi'}$ are 
the global PQ charges of $\phi$ and $\phi'$ with $Q_{\phi}/Q_{\phi'} \neq -10$, respectively.

The global PQ symmetry is broken by the QCD anomaly.
In fact, under the global PQ rotation with a rotation angle $\alpha_{PQ}$,
\begin{eqnarray}
\label{eq:GPQdef}
 \phi \phi'^{10} \to   e^{i\alpha_{PQ}}\times\phi \phi'^{10} \ ,
\end{eqnarray}
the Lagrangian shifts by,
\begin{eqnarray}
\label{eq:anom1}
 {\delta}{\cal L}_{\cancel {PQ}} = \frac{\alpha_{PQ} g_a^2}{32\pi^2} F^a \tilde F^a\ .
\end{eqnarray}
It should be noted that the normalization factor of Eq.\,(\ref{eq:anom1}) is independent of 
the choice of the global PQ charge assignment for the individual fields.

Since the global PQ symmetry is just an accidental one, it is also broken by the Planck suppressed operators explicitly.
However, due to the gauged fiveness symmetry, no PQ-symmetry breaking operators such as $\phi^n$ or $\phi'^n$ ($n>0$) are allowed.
As a result, the explicit breaking terms of the global PQ symmetry are highly suppressed, and 
the lowest dimensional ones are given by,
\begin{eqnarray}
\label{eq:PQbreaking}
{\cal L}_{\cancel{PQ}} \sim \frac{1}{10!} \frac{\phi \phi'^{10}}{M_{\rm PL}^7} + h.c.\ ,
\end{eqnarray}
where  $M_{PL}\simeq2.44\times 10^{18}$ is the reduced Planck scale. 
As we will see in the next section, the breaking terms are acceptably small not to spoil the PQ mechanism 
in a certain parameter space.

%%%%%%%%%%%%%%%%%%%%%%%%%%%%%%%%%%%%%%%%%%
\section{Axion and Global PQ Symmetry}
\label{sec:axion}
To see the properties of the accidental global PQ symmetry,
let us decompose the axion from the would-be Goldstone boson of fiveness.
Both of them originate from the phase components of $\phi$ and $\phi'$;
\begin{eqnarray}
\label{eq:axialcomp}
\phi = \frac{1}{\sqrt{2}} f_a\, e^{i  a/f_a}\ , \quad 
\phi' = \frac{1}{\sqrt{2}}f_b\, e^{i b/f_b}\ ,
\end{eqnarray}
where $f_{a,b}$ are the decay constants and we keep only the Goldstone modes, $a$ and $b$.
The domains of the phase components are given 
\begin{eqnarray}
\label{eq:domain}
\theta_a \equiv {a}/f_a = [0, 2\pi)\ ,  \quad \theta_b\equiv {b}/f_b = [0, 2\pi)\ ,  
\end{eqnarray}
respectively.

In terms of $\theta_{a,b}$, fiveness is realized by,
\begin{eqnarray}
\theta_{a,b} &\to& \theta_{a,b} + q_{a,b} \alpha(x) \\
Y_\mu &\to& Y_\mu  - \partial_\mu \alpha(x)/g  \ .
\end{eqnarray}
Here, $\alpha(x)$ denotes a gauge parameter field with $q_a = -10$ and $q_b = +1$, 
 $Y_{\mu}$ the gauge field, and $g$ the coupling constant, respectively. 
The gauge invariant effective Lagrangian of the Goldstone modes
is given by,
\begin{eqnarray}
{\cal L} &=& \frac{1}{2} f_a^2 D_a^\mu D_{a\mu} +\frac{1}{2} f_b^2 D_b^\mu D_{b\mu}\ ,
\end{eqnarray}
where the covariant derivatives are defined by
\begin{eqnarray}
D_{i}{}_\mu  &=& \partial_\mu \theta_{i} + g q_{i} Y_\mu\  , \,( i = a, b)\ .
\end{eqnarray}

The gauge invariant axion, $A$ ($\propto q_b \theta_a - q_a \theta_b$), and the would-be Goldstone mode, $B$,
are given by
\begin{eqnarray}
\left(
\begin{array}{cc}
A   \\
B
\end{array}
\label{eq:decomp}
\right)=
\frac{1}{\sqrt{q_a^2 f_a^2 + q_b^2f_b^2 }}
\!\!
\left(
\begin{array}{cc}
q_b f_b   &  -q_a f_a   \\
q_a f_a  &   q_b f_b
\end{array}
\right)
\!\!
\left(
\begin{array}{cc}
 a   \\
 b
\end{array}
\right)\ .
\end{eqnarray}
By using $A$ and $B$, the effective Lagrangian is reduced to,
\begin{eqnarray}
{\cal L} &=&  \frac{1}{2}(\q  A)^2 + \frac{1}{2}m_Y^2 \left(Y_\mu - \frac{1}{m_Y}\q_\mu  B \right)^2\ . 
\end{eqnarray}
The second term is the St\"uckelberg mass term of the gauge boson with $m_Y$  being the gauge boson mass,
\begin{eqnarray}
m_Y^2 = g^2(q_a^2 f_a^2 + q_b^2f_b^2 ) \ .
\end{eqnarray}
Through the mass term, the would-be Goldstone mode $B$ is absorbed into $Y_\mu$ by the Higgs mechanism.
The effective decay constant of the axion $A$ is given by,
\begin{eqnarray}
\label{eq:FA}
F_A  = \frac{f_a f_b}{\sqrt{q_a^2 f_a^2 + q_b^2f_b^2 }}\ .
\end{eqnarray}
Given $F_A$, the domain of the gauge invariant axion is given by
\begin{eqnarray}
\frac{A}{F_A} = [0,2\pi)\ .
\end{eqnarray}
when $|q_a|$ and $|q_b|$ are relatively prime integers~\cite{Fukuda:2017ylt}. 

The global PQ symmetry  defined in the previous section is realized by a shift of
\begin{eqnarray}
\label{eq:A}
\frac{A}{F_A}  \to \frac{A}{F_A}  + \alpha_{PQ} \ ,
\end{eqnarray}
where $\alpha_{PQ}$ ranges from $0$ to $2\pi$.
In fact, the phase of the gauge invariant combination $\phi\phi'^{10}$ rotates by
\begin{eqnarray}
 \phi\phi'^{10} \propto e^{iA/F_A} \to e^{i\alpha_{PQ}}e^{iA/F_A} \ ,
\end{eqnarray}
as in Eq.\,(\ref{eq:GPQdef}).

%
%
%, which is accompanied by a global fiveness transformation,
%\begin{eqnarray}
%\label{eq:B}
%\frac{g B}{m_Y} =  \frac{q_b}{q_a} \frac{F_A^2}{f_a^2} \alpha_{PQ}\ .
%\end{eqnarray}
%Here, all the fields with the fiveness charge $q_5$ other than $\phi$ and $\phi'$ have been redefined 
%by multiplying $e^{i g q_5 B/m_Y} $
%so that they are gauge invariant.
%The shifts of $A$ and $B$ induces 
%\begin{eqnarray}
%\phi \to \phi \ , \quad \phi' \to e^{i\alpha_{PQ}/q_a} \phi' \ ,
%\end{eqnarray}
%and hence, the global PQ symmetry in Eq.\,(\ref{eq:globalPQ}) is achieved by Eqs.\,(\ref{eq:A}) and (\ref{eq:B}) with%
%\footnote{Due to the gauge redundancy, ($\phi$, $e^{i2\pi/q_a}\phi'$) is equivalent  to ($\phi$, $\phi'$),
%and hence, the global PQ symmetry is closed in $\alpha_{PQ} = [0,2\pi)$.}
%\begin{eqnarray}
%\label{eq:globalPQ2}
%\mathbf{ {5}}_K' \to e^{i \alpha_{PQ}/q_a}\mathbf{ {5}}_K'\ ,
%\end{eqnarray}
%which is broken by the anomalies.

After integrating out the extra multiplets, the axion obtains anomalous couplings to the SM gauge fields,
\begin{eqnarray}
{\cal L} &=& \frac{g_a^2}{32\pi^2}(N_a\theta_a + N_b \theta_b ) F^a \tilde F^a \cr
&=& \frac{g_a^2}{32\pi^2}(q_b\theta_a - q_a \theta_b ) F^a \tilde F^a
\ ,
\end{eqnarray}
Here, we have used the fact that the numbers of extra multiplets coupling to $\phi$ and $\phi'$ are giving by 
$N_a = q_b = 1$ and $N_b = -q_a = 10$.
By substituting Eq.\,(\ref{eq:decomp}), the anomalous coupling is reduce to,
\begin{eqnarray}
\label{eq:anomA}
{\cal L}  = 
\frac{g_a^2}{32\pi^2}\frac{A}{F_A} F^a \tilde F^a\ ,
\end{eqnarray}
which reproduces the axial anomaly of Eq.\,(\ref{eq:anom1}) by the shift of the axion in Eq.\,(\ref{eq:A}).
Through this term, the axion obtains a mass from the anomalous coupling below the QCD scale, 
with which the QCD vacuum angle is erased.

In the presence of the explicit breaking terms in Eq.\,(\ref{eq:PQbreaking}),
the QCD vacuum angle is slightly shifted by%
\footnote{Hereafter, $\vev{\phi}$ and $\vev{\phi'}$ denote the absolute values of the VEVs of $\phi$ and $\phi'$.}
\begin{eqnarray}
\label{eq:delta1}
{\mit \D}\theta &\sim& 2  \frac{1}{10!} \frac{\vev{\phi} \vev{\phi'}^{10}}{M_{\rm PL}^7 m_a^2 F_A^2} 
\sim 3\times10^{-11} \left(\frac{\vev{\phi}}{10^{10}\,\rm GeV}\right)\left(\frac{\vev{\phi'}}{10^{11}\,\rm GeV}\right)^{10} \ .
\end{eqnarray}
where $m_a$ denotes the axion mass.
Such a small shift should be consistent with the current experimental upper limit on the $\theta$ angle, $\theta \lesssim 10^{-10}$~\cite{Baker:2006ts}.

%%%%%%%%%%%%%%%%%%%%%%%%%%%%
\begin{figure}[t]
\begin{center}
  \includegraphics[width=.4\linewidth]{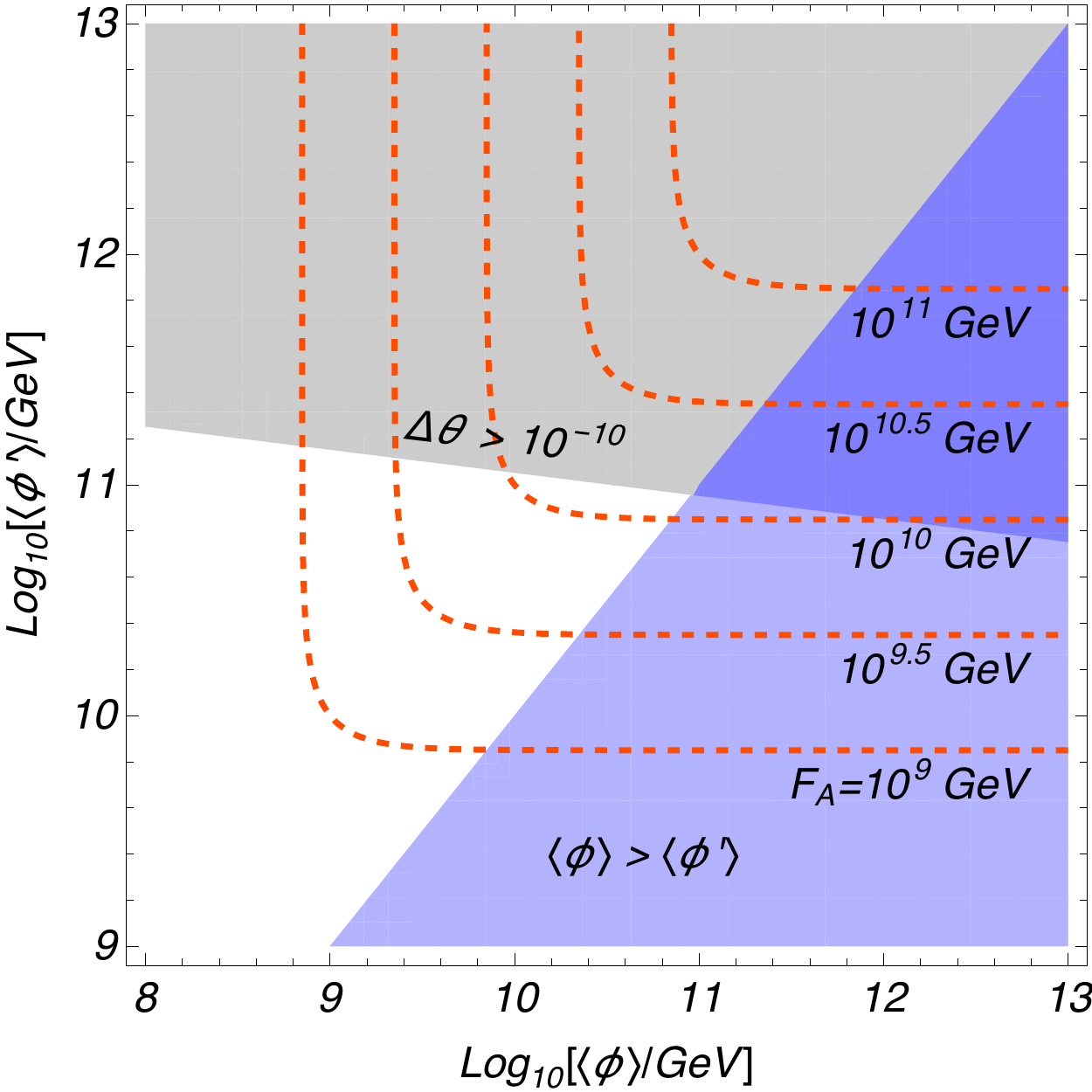}
 \end{center}
\caption{\sl \small
The constraint on the VEVs of $\phi$ and $\phi'$.
The gray shaded region is excluded by ${\mit \D}\theta < 10^{-10}$ for the non-SUSY model (see Eq.\,(\ref{eq:delta1})).
The orange lines are the contours of the effective decay constant $F_A$.
In the blue shaded region, $\vev{\phi} >\vev{\phi'}$.
}
\label{fig:fiveness}
\end{figure}
%%%%%%%%%%%%%%%%%%%%%%%%%%%

In Fig.\,\ref{fig:fiveness}, we show the constraint on the VEVs of $\phi$ and $\phi'$ from
the experimental upper limit on ${\mit\D} \theta$.
In the gray shaded region, the explicit breaking effect  in Eq.\,(\ref{eq:delta1})
is too large to be consistent with ${\mit \D}\theta\lesssim 10^{-10}$.
The orange lines show the contours of the effective decay constant in Eq.\,(\ref{eq:FA}), which is mainly determined by the smaller one 
between $\vev{\phi}$ and $\vev{\phi'}$.
The figure shows that the model is consistent with the the experimental upper limit on ${\mit\D} \theta$ for $\vev{\phi'}\lesssim 10^{11}$\,GeV.
As a result, we find that the gauged PQ mechanism based on the fiveness can solve the strong CP problem 
while satisfying the astrophysical constraint from the observation of supernova 1987A, $F_A \gtrsim 10^8$\,GeV~\cite{Chang:2018rso},
and the condition for successful thermal leptogenesis, 
$M_N = y_N \vev{\phi}\gtrsim 10^9$\,GeV~\cite{Giudice:2003jh,Buchmuller:2005eh,Davidson:2008bu}.	

%%%%%%%
Several  comments are in order.
Since ($\mathbf{\bar 5}_{\rm SM}$,$\mathbf{\bar 5}_{K}$,$\mathbf{\bar 5}_{K}'$) have identical gauge charges,
they are  indistinguishable from each other.
Once $\phi$ and $\phi'$ obtain VEVs in the intermediate scale, $11$-flavors of them become mass partners of ${\mathbf 5}$'s,
and $3$-flavors of them remain massless.
The SM 3-flavors of ${\mathbf {\bar 5}}$ are identified with those massless $\mathbf {\bar 5}$'s.

It should also be noted that the ``inter-sector" interactions via $\mathbf{\bar 5}$ do not lead to explicit breaking of the global PQ
symmetry.
To see this, it is most convenient to choose $Q_{\phi} = 0$ and $Q_{\phi'} = 1$ (see Eq.\,(\ref{eq:PQQ})),
which leads to the global PQ charges,
\begin{eqnarray}
\label{eq:globalPQ}
\phi'(+1)\ , \quad \mathbf{{5}}_K'(+1)\ ,
\end{eqnarray}
with the charges of $\{{\mbox{SM}}, \bar{N}_R,\phi, {\mathbf 5}_K, {\mathbf {\bar 5}}\}$ vanishing.
As the fiveness invariant interactions of ${\mathbf {\bar 5}}$ in Eq.\,(\ref{eq:Lagrangian}) 
are also invariant under the global PQ symmetry in Eq.\,(\ref{eq:globalPQ}),
no explicit breaking terms are generated from the  ``inter-sector"  interactions.%
\footnote{Note that $\phi \phi'^{10}$ is the lowest dimensional operators among all the global PQ breaking operators.
In this case, no larger explicit breaking terms are generated by radiative corrections
other than the anomalous breaking terms given in Eq.\,(\ref{eq:anomA}).}

In the low energy effective theory, the axion couplings to the SM fields are the same with those in the KSVZ axion model except for those to the neutrinos.
As $B-L$ is  an accidental symmetry of the SM except for the neutrino masses, 
the current couplings to the axion proportional to the fiveness can be absorbed by the $B-L$ rotation and $U(1)_Y$ rotation.
The non-vanishing couplings to the neutrinos can also be understood from the fact that the axion in the present model also plays a role of the 
Majoron~\cite{Chikashige:1980ui} which is obvious in the limit of $\vev{\phi'} \gg \vev{\phi}$.
However, it seems very difficult to test the direct couplings between the axion and the neutrinos in laboratory experiments.

\section{Domain Wall Problem}
\label{sec:DW}
Here, let us briefly discuss the domain wall problem and axion dark matter. 
As discussed in \cite{Fukuda:2018oco}, the model suffers from the domain wall problem for $\vev{\phi}\gg \vev{\phi'}$ 
when global PQ symmetry breaking takes place after inflation.
To avoid the domain wall problem, we assume either one of the following possibilities; 
\begin{enumerate}[label=(\roman*)]
\item Both phase transitions of $\vev{\phi}\neq 0$ and $\vev{\phi'}\neq 0$  take place before inflation. 
\item The phase transition, $\vev{\phi'} \neq 0$, takes place before inflation while the transition, $\vev{\phi} \neq 0 $, 
occurs after inflation.
\end{enumerate}
The latter possibility is available as the fiveness charges of $\phi$ and $\phi'$ are 
relatively prime and $|q_a|:|q_b| = 10 : 1$.%
\footnote{The domain wall problem might also be solved for $\vev{\phi}\sim \vev{\phi'}$ 
even if both the phase transitions take place after inflation.
To confirm this possibility, detailed numerical simulations are required.
}

For the first possibility, the cosmic axion abundance is given by,
\begin{eqnarray}
\label{eq:relic}
\Omega_a h^2 \simeq 0.18\,\theta_a^2\left(\frac{F_A}{10^{12}\,{\rm GeV}}\right)^{1.19}\ ,
\end{eqnarray}
for the initial misalignment angle $\theta_a = {\cal O}(1)$~\cite{Lyth:1991ub}. 
Thus, in the allowed parameter region in Fig.\,\ref{fig:fiveness}, i.e., $F_A \lesssim 10^{10}$\,GeV, 
relic axion abundance is a subdominant component of dark matter.
It should be also noted that the Hubble constant during inflation is required to satisfy, 
\begin{eqnarray}
H_{\rm inf} \lesssim 10^{8}\,{\rm GeV} \times \theta_a^{-1}\left(\frac{F_A}{10^{10}\,\rm GeV}\right)^{-0.19}\ .
\end{eqnarray}
to avoid the axion isocurvature problem (see Refs.~\cite{Kawasaki:2013ae,Kawasaki:2018qwp}).%
\footnote{Here, we do not assume that the axion is the dominant component of dark matter but 
use the axion relic abundance in Eq.\,(\ref{eq:relic}) to derive the constraint.
}

For the second possibility, 
the cosmic axion abundance is dominated by the one from the decay of the string-domain wall networks~\cite{Hiramatsu:2012gg} (see Refs.~\cite{Borsanyi:2016ksw,Ballesteros:2016xej,Ringwald:2018xlf} for more recent up-to-date version),
\begin{eqnarray}
\label{eq:axionDM2}
\Omega_a h^2 \simeq 0.035\pm0.012\,\left(\frac{F_A}{10^{10}\,{\rm GeV}}\right)^{1.19}\ .
\end{eqnarray}
Thus, the relic axion from the string-domain wall network can be the dominant component of dark matter
at the corner of the parameter space in Fig.\,\ref{fig:fiveness}.
To avoid symmetry restoration after inflation, we also require that the maximum temperature during reheating~\cite{Kolb:1990vq}, 
\begin{eqnarray}
T_{\rm MAX}  \simeq g_*^{-1/8}T_R^{1/2}H_{\rm inf}^{1/4} M_{PL}^{1/4} \ ,
\end{eqnarray}
does not exceed $\vev{\phi'}$, which leads to
\begin{eqnarray}
\label{eq:HinfUP}
H_{\rm inf} \lesssim 5\times 10^{8}\,{\rm GeV}  \left(\frac{\vev{\phi'}}{10^{11}\,\rm GeV}\right)^{4}\left(\frac{10^{9}\,\rm GeV}{T_R}\right)^{2}\ .
\end{eqnarray}
Here, we use the effective massless degrees of freedom $g_* \simeq 200$, though the condition does not depend on $g_*$ significantly.

%%%%%%%%%%%%%%%%%%%%%%%%%%%%%%%%%%%%%%%%
\section{Supersymmetric Extension}
\label{sec:susy}
The SUSY extension of the present model is straightforward.
The SM matter fields, the right-handed neutrinos, and the extra multiplets are simply extended to corresponding supermultiplets
with the same fiveness charges given in Eqs.\,(\ref{eq:SMfiveness}) and (\ref{eq:charge}).
The Higgs doublets are extended to the two Higgs doublet supermultiplets $H_u$ and $H_d$ as in the minimal SUSY Standard Model (MSSM).
The fiveness charges are assigned to be $H_u(-2)$ and $H_d(+2)$, respectively.
The complex scalars $\phi$ and $\phi'$ are also extended to corresponding supermultiplets  which are accompanied 
by supermultiplets with opposite fiveness charges, $\bar{\phi}$ and $\bar{\phi}'$ (see Tab.\,\ref{tab:R}).

Under the fiveness symmetry, the superpotential is restricted to%
\footnote{More generally, the Higgs bi-linear, $H_u H_d$, also couples to $X$ and $Y$.
We assume that the soft masses of the Higgs doublets are positive and larger than those of $\phi$'s and $\phi'$'s,
so that the Higgs doublets do not obtain VEVs from the couplings to $X$ and $Y$.
We may also restrict those couplings by some symmetry.
} 
\begin{eqnarray}
\label{eq:W}
W &=&  {\mathbf{10}_{\rm SM}}  {\mathbf{10}_{\rm SM}} H_u
+{\mathbf{10}_{\rm SM}}  {\mathbf{\bar{5}}} H_d
+ {\mathbf{\bar{5}}} \bar{N}_R H_u
- \frac{1}{2} \phi\, \bar{N}_{R}\bar{N}_{R} 
+ \bar\phi \, \mathbf{5}_K \mathbf{\bar 5}+ \bar\phi' \, \mathbf{5}_K' \mathbf{\bar 5}\cr
&&+ X(2\phi\bar\phi -v^2) + Y(2\phi' \bar\phi' - v'^2) \ .
\end{eqnarray}
%{\bf HOW CAN WE DISTINGUISH $H_uH_d$ from $\phi \bar{\phi}$?}\\
Here, $X$ and $Y$ are introduced to make $\phi$ and $\phi'$ obtain non-vanishing VEVs, which 
are neutral under fiveness.%
\footnote{See \cite{Fukuda:2018oco} for details of the SUSY extension of the gauged PQ mechanism.} 
The coupling coefficients are again omitted for notational simplicity.
The SUSY extension again possesses the global PQ symmetry as in the case of the non-SUSY model.

%%%%%%%%%%%%%%%%%%%%%%%
\begin{table}[t]
\caption{The charge assignment of the fiveness symmetry and the gauged ${\mathbb Z}_{4R}$ symmetry.
Here, we fix the ${\mathbb Z}_{4R}$ charges of the Higgs doublets to $0$ which is motivated by
pure gravity mediation model~\cite{Ibe:2006de,*Ibe:2011aa,*Ibe:2012hu}. 
An extra multiplet $({\mathbf 5}_E, {\mathbf {\bar 5}}_E)$ is introduced to cancel the ${\mathbb Z}_{4R}$--$SU(5)^2$ anomaly~\cite{Kurosawa:2001iq}.
}
\small{
\begin{center}
\begin{tabular}{|c||Y|Y|Y|Y|Y|Y|Y|Y|Y|Y|Y|Y|Y||Y|Y|}
\hline
 & ${\mathbf{10}}_{\rm SM}$&${\mathbf {\bar 5}}$ & $\bar{N}_R$ & $H_u$ & $H_d$ &$\mathbf{5}_K$ &${\mathbf 5}_K'$ & $\phi$ &$\bar\phi$ & $\phi'$ & $\bar\phi'$ & $X$ &$Y$& $\mathbf 5_E$& $\mathbf{ \bar 5}_E$
  \\ \hline
$\mbox{fiveness}$ & $+1$ &$-3$ &$+5$ &$-2$ &$+2$  & $-7$ &$+4$ & $-10$ & $+10$ & $+1$ & $-1$& $0$ & $0$& $+3$ & $-3$
\\ \hline
${\mathbb Z}_{4R}$ & $+1$ &$+1$ &$+1$ &$0$ &$0$  & $+1$ &$+1$ & $0$ & $0$ & $0$& $0$& $+2$& $+2$ & $-1$ & $+1$
\\ \hline
\end{tabular}
\end{center}}
\label{tab:R}
\end{table}%
%%%%%%%%%%%%%%%%%%%%%%%

In addition to fiveness, we also assume that a discrete subgroup of $U(1)_R$, $\mathbb{Z}_{NR}~(N>2)$, 
is an exact discrete gauge symmetry.
This assumption is crucial  to allow the VEV of the superpotential, and hence, the  supersymmetry breaking scale 
much smaller than the Planck scale.%
\footnote{$R$-symmetry  is also relevant for SUSY breaking vacua 
to be stable\,\cite{Affleck:1984xz,Nelson:1993nf}.}
In the following, we take the simplest possibility, ${\mathbb Z}_{4R}$
with the charge assignment given in Tab.\,\ref{tab:R},
which is free from ${\mathbb Z}_{4R}$--$SU(5)^2$ anomaly and the gravitational anomaly.%
\footnote{It should be noticed that there is no need to add extra $SU(5)$ singlet fields to cancel the anomalies.}
It should be noted that the mixed anomalies of ${\mathbb Z}_{4R}$ and fiveness do not put constraints on charges since they depend on the 
normalization of the heavy spectrum~\cite{Krauss:1988zc,Preskill:1990bm,Preskill:1991kd,Banks:1991xj,Ibanez:1991hv,Ibanez:1992ji,Csaki:1997aw,Lee:2010gv,Fallbacher:2011xg,Evans:2011mf}.%
\footnote{GUT models consistent with the ${\mathbb Z}_{4R}$ symmetry are  discussed in, e.g.,~\cite{Izawa:1997he,Harigaya:2015zea}.}

Under fiveness and the gauged ${\mathbb Z}_{4R}$ symmetry, the lowest dimensional operators which break the global PQ symmetry are given by,
\begin{eqnarray}
W_{\cancel{PQ}} =  \frac{m_{3/2}}{10!M_{\rm PL}}\frac{\phi \phi'^{10}}{M_{\rm PL}^8} +  \frac{m_{3/2}}{10!M_{\rm PL}}\frac{\bar{\phi} \bar\phi'^{10}}{M_{\rm PL}^8}\ .
\end{eqnarray}
It should be noted that a lower dimensional PQ breaking term, $\bar{\phi'}^5 {\bar N}_R$, is forbidden by the ${\mathbb Z}_{4R}$ symmetry.
The above superpotential contributes to the shift of the QCD vacuum angle mainly through the scalar potential,
\begin{eqnarray}
\label{eq:PQbreaking2}
{\cal L}_{\cancel{PQ}} \sim \frac{8m_{3/2}^2}{10! M_{\rm PL}} \frac{\phi \phi'^{10}}{M_{\rm PL}^8} +  \frac{8m_{3/2}^2}{10!M_{\rm PL}} \frac{\bar{\phi} \bar\phi'^{10}}{M_{\rm PL}^8} + h.c. \ ,
\end{eqnarray}
where $m_{3/2}$ denotes the gravitino mass.
Compared with Eq.\,(\ref{eq:PQbreaking}), the explicit breaking is suppressed by a factor of $(m_{3/2}/M_{\rm PL})^2$.
Accordingly, the shift of the QCD vacuum angle is given by,
\begin{eqnarray}
\label{eq:delta2}
{\mit \D}\theta &\sim& 2  \frac{1}{10!} \frac{8m_{3/2}^2\vev{\phi} \vev{\phi'}^{10}}{M_{\rm PL}^9 m_a^2 F_A^2} 
\sim 10^{-25} \left( \frac{m_{3/2}}{100\,\rm TeV}\right)^2\left(\frac{\vev{\phi}}{10^{11}\,\rm GeV}\right)\left(\frac{\vev{\phi'}}{10^{12}\,\rm GeV}\right)^{10}
\ ,
\end{eqnarray}
where we assume $\vev{\phi} = \vev{\bar\phi}$ and $\vev{\phi'} = \vev{\bar\phi'}$ for simplicity.%
\footnote{The following argument can be easily extended to the cases with $\vev{\phi}\neq \vev{\bar\phi}$ and $\vev{\phi'} \neq \vev{\bar\phi'}$. }

%%%%%%%%%%%%%%%%%%%%%%%%%%%%
\begin{figure}[t]
 \begin{center}
  \includegraphics[width=.4\linewidth]{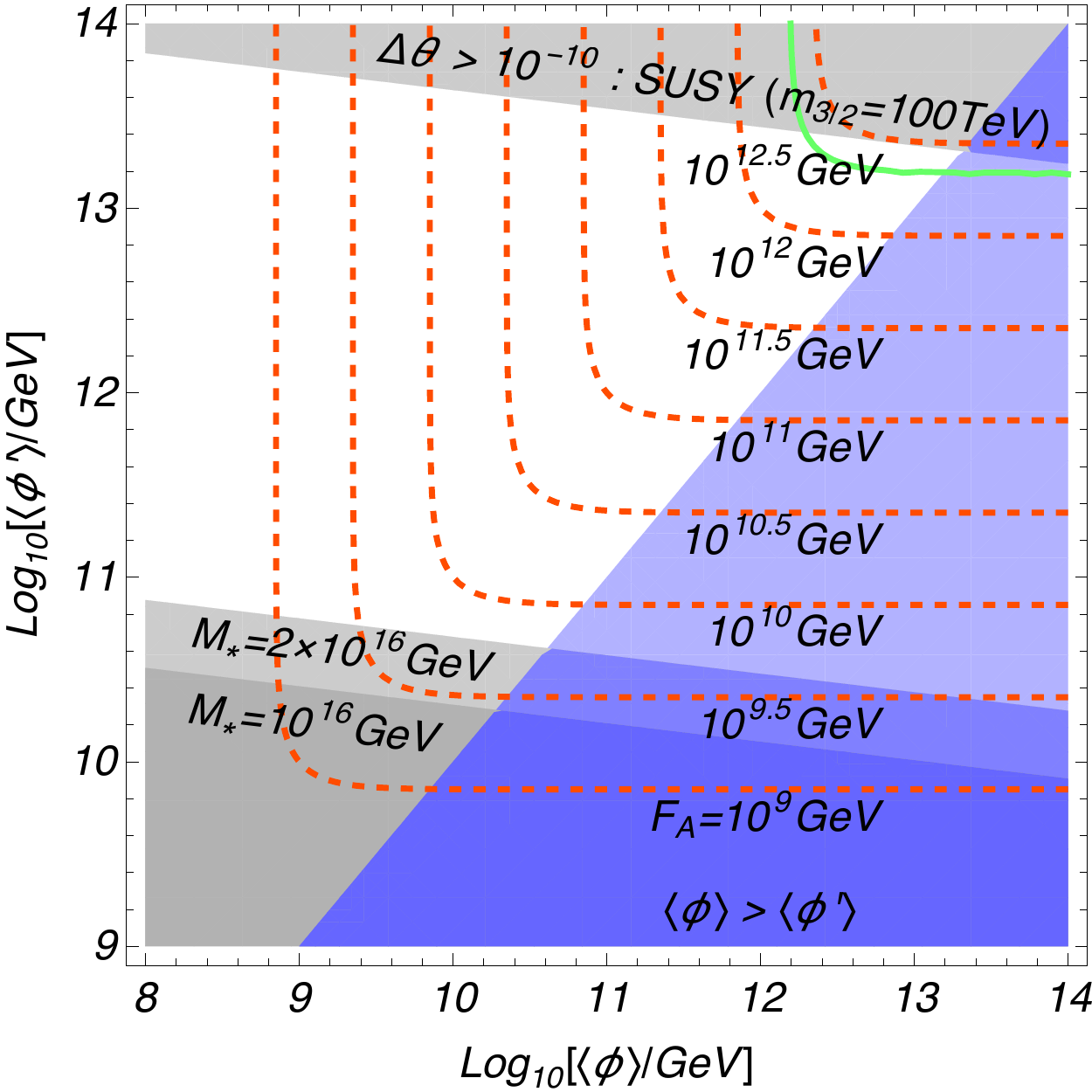}
 \end{center}
\caption{\sl \small
The constraint on the VEVs of $\phi$ and $\phi'$ for the SUSY extension.
The gray shaded upper region is excluded for the SUSY model with $m_{3/2} = 100\,{\rm TeV}$ (see Eq.\,(\ref{eq:delta2})).
The orange lines are the contours of the effective decay constant $F_A$.
In the blue shaded region, $\vev{\phi} > \vev{\phi'}$.
The gray shaded lower regions are excluded as the gauge coupling constants become non-perturbative below the GUT scale. 
The thin green region is excluded by the Axion Dark Matter eXperiment (ADMX) \cite{Du:2018uak}
where the dark matter density is assumed to be dominated by the relic axion.
}
\label{fig:fivenessSUSY}
\end{figure}
%%%%%%%%%%%%%%%%%%%%%%%%%%%

In Fig.\,\ref{fig:fivenessSUSY}, we show the constraints on the VEVs of $\phi$ and ${\phi'}$ from
the experimental upper limit on ${\mit\D} \theta$.
Here, we take the gravitino mass, $m_{3/2}\simeq 100$\,TeV,
which is favored to avoid the cosmological gravitino problem
for $T_R \gtrsim 10^9$\,GeV~\cite{Kawasaki:2004qu,Jedamzik:2006xz,Kawasaki:2017bqm}. 
For $m_{3/2}\simeq 100$\,TeV,  the scalar partner and the fermionic partner of the axion  also do not cause cosmological problems
as they obtain the masses of the order of the gravitino mass and decay rather fast~\cite{Kawasaki:2007mk}.
In the figure, the gray shaded region is excluded by the constraint on ${\mit \D}\theta\lesssim 10^{-10}$.
Due to the suppression of the breaking term in Eq.\,(\ref{eq:PQbreaking2}), the higher value of $\vev{\phi'}$
is allowed compared with the non-SUSY model.
The higher $\vev{\phi'}$ is advantageous to avoid symmetry restoration after inflation (see Eq.\,(\ref{eq:HinfUP})),
with which the domain wall problem is avoided in the possibility (ii) (see section\,\ref{sec:axion}).
Accordingly, the decay constant can also be as high as about $10^{11-12}$\,GeV, 
which also allows the axion to be the dominant dark matter component (see Eq.\,(\ref{eq:axionDM2})).
Therefore, we find that the SUSY extension of the model is more successful.%
\footnote{As in \cite{Fukuda:2018oco}, we will discuss a possibility where SUSY and $B-L$ are broken simultaneously elsewhere.}

It should be noted that the $11$-flavors of extra multiplets  at the intermediate scale make the renormalization group running of the MSSM gauge coupling constants asymptotic non-free. 
Thus, the masses of them are bounded from below so that  perturbative unification is achieved.
In the figure, the gray shaded lower region shows the contour of the renormalization scale $M_*$ at which at least one of $g_{1,2,3}$ becomes $4\pi$.
Here, we use the one-loop renormalization group equations assuming that the extra quarks obtain masses of $\vev{\phi}$ and $\vev{\phi'}$, respectively.%
\footnote{The  masses of the sfermions, the heavy charged/neutral Higgs boson, the Higgsinos, and $({\mathbf 5}_E, {\mathbf{\bar 5}}_E)$ 
 are at the gravitino mass scale,
$m_{3/2}\simeq 100$--$1000$\,TeV.
The gaugino masses are, on the other hand, assumed to be in the TeV scale as expected by anomaly mediation~\cite{Giudice:1998xp,Randall:1998uk}. 
This is motivated by the pure gravity mediation model in\,\cite{Ibe:2006de,*Ibe:2011aa,*Ibe:2012hu} 
(see also Refs.\,\cite{Giudice:2004tc,ArkaniHamed:2004yi,Wells:2004di,ArkaniHamed:2012gw} 
for similar models), where the Higgsino mass is generated from the $R$-symmetry breaking~\cite{Casas:1992mk}.}
The result shows that the perturbative unification can be easily achieved for $\vev{\phi'} \gtrsim 10^{9\mbox{--}10}$\,GeV
even in the presence of $11$-flavors of the extra multiplets.

\section{Conclusions and Discussions} 
In this paper, we consider the gauged PQ mechanism where the gauged PQ symmetry is identified with the $B-L$ symmetry (fiveness).
As the ${B-L}$ gauge symmetry is the most plausible extension of the SM,
the identification of the gauged PQ symmetry with  $B-L$ is very attractive.
An intriguing coincidence between  the $B-L$ breaking scale appropriate for the thermal leptogenesis and the favored PQ breaking scale 
from the astrophysical constraints also motivates this identification.

We found a natural $B-L$ charge assignment motivated by the seesaw mechanism in the $SU(5)$ GUT,
with which the gauged PQ mechanism is achieved.
There, the global PQ symmetry breaking effects are suppressed by the gauged fiveness symmetry 
so that the successful PQ mechanism is realized.
As a notable feature, the fiveness charge assignment does not require extra $SU(5)$ singlet matter fields other than the right-handed neutrinos
to cancel the $[U(1)_{gPQ}]^3$ anomaly and the gravitational anomaly.
This feature is advantageous since the singlet fields required for anomaly cancellation tend to be rather light and longlived,
and hence, often cause cosmological problems.
As a result, we find that the gauged PQ mechanism based on the $B-L$ symmetry is successfully consistent with thermal leptogenesis. 

We also discussed the SUSY extension  where the ${\mathbb Z}_{4R}$ symmetry is also assumed.
As has been shown, a larger effective decay constant is allowed in the SUSY model, as explicit breaking of the global PQ symmetry is more suppressed.
Resultantly, the upper limit on the effective decay constant is extended to
\begin{eqnarray}
F_A \lesssim 10^{12.5}\,{\rm GeV}\ ,
\end{eqnarray}
which corresponds to the axion mass,
\begin{eqnarray}
m_a \gtrsim 1.9\,{\mu \rm eV} 
\left(
\frac{10^{12.5}\,\rm GeV}{F_A}
\right)\ .
\end{eqnarray}
The dark matter axion in this mass range can be detected by the  ongoing ADMX-G2 experiment~\cite{Asztalos:2009yp}
and future ADMX-HF experiment~\cite{vanBibber:2013ssa}.

In the SUSY model, it should be also noted  that  ${\mathbb Z}_{4R}$ is spontaneously broken down to the ${\mathbb Z}_{2R}$ symmetry.%
\footnote{For cosmological implication of the spontaneous discrete $R$-symmetry breaking, see Ref.~\cite{Harigaya:2015yla}.}
Thus, the lightest supersymmetric particle in the MSSM sector also contributes to the dark matter density.
Therefore, the model predicts a wide range of dark matter scenario from  axion dominated dark matter 
to the LSP dominated dark matter, which can be tested by future extensive dark matter searches.

As emphasized above, the fiveness anomalies are canceled without introducing singlet fields other than the right-handed neutrinos.
Although this feature is advantageous from the cosmological point of view, the fundamental reason for the cancellation remains an open question.
In the case of the SM and the right-handed neutrino sector, the anomaly cancellation of the fiveness can be 
explained by the $SO(10)$ unification in which the fiveness becomes a part of the $SO(10)$ gauge symmetry.
In the present model, however, it cannot be unified into the $SO(10)$ or larger unified groups simply.
At this point, the anomaly cancellation is a mere accident, and we have not found any deeper insights on the anomaly cancellation
(see discussion in the appendix~\ref{sec:uniqueness}).

%%%%%%%%%%%%%%%%%%%%%%%%%%%%%%%%%%%%%%%
%%%%%%%%%%% Acknowledgments %%%%%%%%%%%
%%%%%%%%%%%%%%%%%%%%%%%%%%%%%%%%%%%%%%%
\begin{acknowledgments}
The authors would like to thank H.~Fukuda for valuable discussion at the early stage of this work.
This work is supported in part by Grants-in-Aid for Scientific Research from the Ministry of Education, Culture, Sports, Science, and Technology (MEXT) 
KAKENHI, Japan,  No.\,15H05889 and No.\,16H03991 (M. I.),  No.\,26104001 and No.\,26104009 and No.\,16H02176  (T. T. Y.); 17H02878 (M. I. and T. T. Y.), 
and by the World Premier International Research Center Initiative (WPI), MEXT, Japan.
The work of M. S. is supported in part by a Research Fellowship for Young Scientists from the Japan Society for the Promotion of Science (JSPS).
\end{acknowledgments}

\appendix
\section{Uniqueness of the Fiveness Charge}
\label{sec:uniqueness}
In this appendix, we discuss the uniqueness of the charge assignment in Eq.\,(\ref{eq:charge}), which might help 
to understand the origin of the anomaly free fiveness.
First, let us suppose that there are ($n+3$)-flavors of ${\mathbf {\bar 5}}$ fermions with the fiveness charge $-3$.%
\footnote{The choice of $-3$ just defines the normalization of fiveness.}
For $n=0$, anomaly free fiveness and $SU(5)$ gauge symmetries are achieved by introducing $3$-flavors of ${\mathbf {10}}_{\rm SM}$  and $\bar{N}_R$
with the fiveness charges $+1$ and $+5$, respectively while allowing the first four Yukawa interactions in Eq.\,(\ref{eq:Lagrangian}).
For $n>0$, on the other hand, the anomaly free conditions require more fermions.
Given the fact that the SM consists of $3$ flavors, it is simplest to add $n$-flavors of ${\mathbf 5}$ fermions to cancel self- and gravitational anomalies 
of $SU(5)$.
When a $\mathbf 5$ fermion has fiveness charge $+3$, it becomes a mass partner of one of the $n$-flavors of ${\mathbf {\bar 5}}$,
which  ends up with a model with ($n-1$)-flavors of ${\mathbf {\bar 5}}$.
Thus, we assume that the charge of $\mathbf 5$'s are not equal to $+3$.

The anomaly free charge assignment of $\mathbf 5$'s is fixed in the following way.
For all the $n$-flavors of (${\mathbf {5}}$, ${\mathbf {\bar 5}}$) to have masses in the intermediate scale,
they need to couple to the order parameters of fiveness.
As we assume the seesaw mechanism, we have a natural candidate of such an order parameter, 
a complex scalar field, $\phi$, with a fiveness charge $-10$.
In order to make all the  $n$-flavors of (${\mathbf {5}}$, ${\mathbf {\bar 5}}$) massive while achieving anomaly free fiveness, 
however, it is required to introduce one more complex scalar, $\phi'$, with the fiveness charge $q_{\phi'}$.

In the presence of $\phi$ and $\phi'$, the mass terms of  (${\mathbf {5}}$, ${\mathbf {\bar 5}}$) are generated from
\begin{eqnarray}
{\cal L} = \phi \,{\mathbf {5}}\,{\mathbf {\bar 5}} +  \phi^* \,{\mathbf {5}}'\,{\mathbf {\bar 5}}
+ \phi' \,{\mathbf {5}}''\,{\mathbf {\bar 5}} +  \phi'^* \,{\mathbf {5}}'''\,{\mathbf {\bar 5}} \ ,
\end{eqnarray}
where the coupling coefficients are again omitted.
Here,  ${\mathbf 5}$'s are devided into \{${\mathbf 5}$,${\mathbf 5}'$,${\mathbf 5}''$,${\mathbf 5}'''$\} whose fiveness charges are given by,
\begin{eqnarray}
{\mathbf 5}(+13)\ , \quad {\mathbf 5}'(-7)\ , \quad {\mathbf 5}''(-q_{\phi'}+3)\ , \quad {\mathbf 5}'''( q_{\phi'} + 3)\ ,
\end{eqnarray}
respectively. 
We allocate $N_5$, $N_5'$, $N_5''$ and $N_5'''$ flavors to \{${\mathbf 5}$,${\mathbf 5}'$,${\mathbf 5}''$,${\mathbf 5}'''$\} with 
$N_5 + N_5' + N_5''+N_5''' = n$.t
The anomaly free conditions of fiveness are given by,
\begin{eqnarray}
13^3 N_5  - 7^3 N_5' +  (-q_{\phi'} + 3)^3N_5'' + (q_{\phi'} +3)^3 N_5''' - 3^3 n&= &0\ , \\
13 N_5  - 7 N_5' +  (-q_{\phi'} + 3)N_5'' + (q_{\phi'} +3) N_5''' - 3 n &=&0\ . 
\end{eqnarray}
By solving the anomaly free conditions, we find only two sets of solutions,
\begin{eqnarray}
 N_5 = 0\ , \quad N_5' = 1\ , \quad N_5'' = 0 \ , \quad N_5''' = 10 \ , \quad q_{\phi'} = 1 \ , 
 \end{eqnarray}
 or
 \begin{eqnarray}
 N_5 = 7\ , \quad N_5' = 1\ , \quad N_5'' = 3 \ , \quad N_5''' = 0 \ , \quad q_{\phi'} = 20 \ , 
 \end{eqnarray}
 both of which corresponds to $n = 11$.%
 \footnote{We take $q_{\phi'} > 0$ without loss of generality.}
Here, we restrict ourselves to $n < 22$.
The first charge assignment is nothing but the fiveness charges assumed in this paper,
while the later is another possibility.
In this sense, we find that the number of the flavors, $n = 11$, is a unique choice within $n < 22$, 
and the fiveness charge assignment in this paper is one of the only two possibilities,
where the second possibility is not suitable for the gauged PQ mechanism.

\bibliography{papers}
%\bibliography{draft_arxiv_2.bbl}

\end{document}